\RequirePackage{ifpdf}
\ifpdf 
\documentclass[pdftex]{sigma}
\else
\documentclass{sigma}
\fi

\def\R{\mathbb{R}}

\begin{document}

\allowdisplaybreaks

\renewcommand{\PaperNumber}{115}

\FirstPageHeading

\renewcommand{\thefootnote}{$\star$}

\ShortArticleName{Second-Order Approximate Symmetries of the
Geodesic Equations}

\ArticleName{Second-Order Approximate Symmetries \\ of the
Geodesic Equations for the Reissner--Nordstr\"om Metric and
Re-Scaling of Energy of a Test Particle\footnote{This paper is a contribution to the Proceedings
of the Seventh International Conference ``Symmetry in Nonlinear
Mathematical Physics'' (June 24--30, 2007, Kyiv, Ukraine). The
full collection is available at
\href{http://www.emis.de/journals/SIGMA/symmetry2007.html}{http://www.emis.de/journals/SIGMA/symmetry2007.html}}}

\Author{Ibrar HUSSAIN~$^\dag$, Fazal M. MAHOMED~$^\ddag$ and Asghar QADIR~$^{\dag}$}

\AuthorNameForHeading{I.~Hussain, F.M.~Mahomed and A.~Qadir}

\Address{$^\dag$~Centre for Advanced Math. and Phys., 
National University of Sciences and Technology,\\
$\phantom{^\dag}$~Campus of the College of Electr. and Mech. Eng., Peshawar Road, Rawalpindi, Pakistan}

\EmailD{\href{mailto:ihussain@camp.edu.pk}{ihussain@camp.edu.pk},
\href{mailto:aqadirmath@yahoo.com}{aqadirmath@yahoo.com}}

\Address{$^\ddag$ 
School of Computational and Applied Mathematics, University of the Witwatersrand,\\
$\phantom{^\ddag}$~Wits 2050, South Africa}
\EmailD{\href{mailto:Fazal.Mahomed@wits.as.za}{Fazal.Mahomed@wits.as.za}}

\ArticleDates{Received August 14, 2007, in f\/inal form November
16, 2007; Published online December 07, 2007}

\Abstract{Following the use of approximate symmetries for the
Schwarzschild spacetime by  A.H.~Kara,  F.M.~Mahomed and A. Qadir ({\it Nonlinear Dynam.}, to appear), we have investigated the exact and
approximate symmetries of the system of geodesic equations for the
Reissner--Nordstr\"om spacetime (RN). For this purpose we are forced
to use second order approximate symmetries. It is shown that in the
second-order approximation, energy must be rescaled for the RN
metric. The implications of this rescaling are discussed.}


\Classification{83C40; 70S10}

\section{Introduction}

Though most of the well known spacetimes of General Relativity (GR)
are time-translational invariant, generally this is not guaranteed
\cite{a18}. Consequently energy is not generally conserved and hence
there is no good def\/inition for energy \cite{a36}. Nowhere is this
problem faced as severely as in gravitational wave spacetimes, where
energy is not given by the stress-energy tensor, which is zero
\cite{a18}. For these spacetimes one can check that they impart
energy to test particles in their path \cite{a37,a38} despite the
fact that the stress-energy tensor is zero. The lack of a good
def\/inition of energy, also leads to problems with the def\/inition of
mass \cite{a36}. As such, one needs to use some concept of time
symmetry that allows for slight deviations away from exact symmetry.

This approach was attempted earlier by various people. There have
been a number of dif\/ferent def\/initions of ``approximate symmetry''.
One idea was to assume that conservation of energy holds
$\it{asymptotically}$ \cite{a21} and to examine whether it would work
for gravitational radiation and to def\/ine a positive def\/inite
energy. This seems unsatisfactory as the gravitational energy should
reach (the place) inf\/inity. There may then be problems with orders
of approximation being consistent. An altogether dif\/ferent approach
was taken by providing a measure of the extent of break-down of
symmetry. The integral of the square of the symmetrized derivative
of a vector f\/ield was divided by its mean square norm
\cite{a22,a23}. This led to what was called an $\textit{almost
symmetric space}$ and the corresponding vector f\/ield an
$\textit{almost Killing vector}$ \cite{a24}. This measure of
``non-symmetry'' in a given direction was applied to the Taub
cosmological solution~\cite{b24} and to study gravitational
radiation. It provides a choice of gauge that makes calculations
simpler and was used for this purpose \cite{c24}. Essentially based
on the almost symmetry, the concept of an ``approximate symmetry
group'' was presented \cite{a25}. However it has not been
unequivocally successful either. The approach of a slightly broken
symmetry seems promising, but merely providing simplicity of
calculations is not physically convincing. Other approaches need to
be tried, to f\/ind one that seems signif\/icantly better than
others.

The development of geometry has been driven by its application to
kinematics and dyna\-mics. As such, one might look at a geometrically
driven def\/inition of symmetry. On the other hand, we are concerned
with conservation laws, which are given by the invariants of the
Euler-Lagrange equations. Hence, for our purposes, Lie symmetries as
embodied in isometries and in Noether's theorem, should provide us
the desired approach to def\/ine ``approximate symmetries". Now, there
is a connection between isometries and the symmetries of
dif\/ferential equations (DEs) through the geodesic equations
\cite{a19,a13}. We propose that the Baikov--Gazizov--Ibragimov concept
of ``approximate symmetry''~\cite{b11,a11} could be extended and
adapted for the purpose of def\/ining energy in GR via ``approximate
isometries'' through the above-mentioned connection between the
symmetries of geometry and the symmetries of dif\/ferential equations.
Using that connection approximate symmetries of the Schwarzschild
metric were discussed~\cite{a1}. The Schwarzschild metric has much
fewer symmetries than the Minkowski metric (only 4~generators). One
would expect that in the limit of small gravitational mass the
``lost'' symmetries should be ``approximately'' recovered. Of course,
this is a static metric and energy is conserved and it is not
relevant for a discussion of energy. However, linear and spin
angular momentum conservation is lost. The ``trivial'' approximate
symmetries of the geodesic equations recover these as approximately
conserved quantities. To be able to interpret the approximate
symmetry results so obtained we need to f\/irst follow how higher
order approximate symmetries are to be understood for our purpose
and then how the lack of time-translational symmetry on a globally
def\/ined spacelike hypersurface is to be interpreted. As such, before
dealing with gravitational waves it is necessary to study the
Reissner--Nordstr\"om (RN) metric and the Kerr metric from this point
of view. In this paper we present the discussion of RN spacetime
leaving the other spacetimes for subsequent consideration.

In going to the RN spacetime the approximate symmetries and
conservation laws, which were recovered in the f\/irst-order
approximation~\cite{a1}, are lost. Nor is there any approximate
symmetry for the reduced orbital equation. However one would expect
that in the limit of small charge we should recover the lost
symmetries. To recover the conservation laws lost again we need to
appeal to second-order approximate symmetries. It makes no
dif\/ference whether the exact or perturbed equation is used in the
def\/inition of f\/irst-order approximate symmetries. In this paper it
is shown that in the def\/inition of second-order approximate
symmetries it makes a signif\/icant dif\/ference. We use the perturbed
equation in the def\/inition of second-order approximate symmetries,
the expression for energy of a test particle in the RN metric is
re-scaled.

The plan of the paper is as follows. In the next section we explain
the concept of symmetry and approximate symmetry of DEs, and
approximate symmetries of the Schwarzschild metric. In Section~3
approximate symmetries of the orbital equation for the RN metric and
of the system of geodesic equations for this metric are discussed.
Finally a summary and discussion is given in Section~4.

\section{Symmetries and approximate symmetries of DEs}

Symmetries are very useful because they are directly connected to
the conservation laws through Noether's theorem~\cite{a2,a3,a4}. If,
for a given system of DEs there is a variational principle, then a
continuous symmetry under which the action functional remains
invariant yields a conservation law \cite{a5,a6,a7,a8}. Here we are
interested in Lie symmetry methods~\cite{a9}. The set of
inf\/initesimal symmetry generators of a system of dif\/ferential
equations form a Lie algebra \cite{a6,a10,a11}. Geometrically the
symmetries of a manifold are characterized by its \textit{Killing
vectors}, or \textit{isometries}, which also form a Lie algebra
\cite{a12}. The symmetries of the manifold are inherited by the
geodesic equations on it with additional symmetries~\cite {a13}.
They give quantities conserved under geodesic motion~\cite{a14} and
lead to f\/irst integrals of the geodesic equations~\cite{a15}.

It may often happen that a manifold does not possess exact symmetry
but approximately does so. We may be able to obtain more information
from near symmetry or broken symmetry, than from the exact symmetry
always maintained. As such approximate symmetries of a manifold are
worth exploring. Methods for obtaining the approximate symmetries of
a system of ordinary dif\/ferential equations (ODEs) are available in
the literature~\cite{a16,b11}.

A system of $p$ ODEs of order $n$ \cite{a10,a11}%
\begin{gather*}
\mathbf{E}\big(s;\mathbf{x}(s),\mathbf{x}^{\prime }(s),\mathbf{x}%
^{\prime \prime }(s),\dots,\mathbf{x}^{(n)}(s)\big)=0,
\end{gather*}%
has the \textit{symmetry generator}%
\begin{gather}
\mathbf{X}^{[n]} = \xi (s,\mathbf{x})\frac{\partial
}{\partial s}\!+\eta ^{\alpha }(s,\mathbf{x})\frac{\partial }{\partial
x^{\alpha }}\!+\eta _{,s}^{\alpha }(s,\mathbf{x},\mathbf{x}^{\prime
})\frac{\partial }{\partial x^{\alpha \prime }}\!+\!\cdots\!+\eta
_{,(n)}^{\alpha }(s,\mathbf{x},\mathbf{x}^{\prime
},\dots,\mathbf{x}^{(n)})\frac{\partial }{\partial x^{\alpha (k)}},
\label{2}
\end{gather}
if $\mathbf{X}^{[n]}$ annihilates the system of equations
$\mathbf{E}$ (for solutions of the equation)
\begin{gather*}
\mathbf{X}^{[n]}\left( \mathbf{E})\right\vert _{\mathbf{E}=0} =0,  
\end{gather*}
where $\mathbf{x}$ is a point in the underlying $n$-dimensional
space and $\mathbf{x}^{(n)}$ is the $n$th derivative with respect to
$s$. The prolongation coef\/f\/icients are given by
\begin{gather*}
\eta _{,(n)}^{\alpha }=\frac{d\eta _{,(n-1)}^{\alpha
}}{ds}-x^{\alpha (n)} \frac{d\xi }{ds},\quad n\geq 1,\qquad {\rm
where}\quad \eta _{,(1)}^{\alpha }=\eta _{,s}^{\alpha }.  
\end{gather*}%
If $p=1$ then the system reduces to a single equation.

The $\mathit{k}$th-order \textit{approximate symmetry} of a
perturbed system of ODEs
\begin{gather}
\mathbf{E}=\mathbf{E}_{0}+\epsilon \mathbf{E}_{1}+\epsilon ^{2}\mathbf{E}%
_{2}+\cdots+\epsilon ^{k}\mathbf{E}_{k}+O(\epsilon ^{k+1}) \label{5}
\end{gather}
is given by the generator
\begin{gather}
\mathbf{X}=\mathbf{X}_{0}+\epsilon \mathbf{X}_{1}\mathbf{+}\epsilon
^{2}\mathbf{X}_{2}+\cdots +\epsilon ^k\mathbf{X}_k, \label{6}
\end{gather}
if the symmetry condition
\begin{gather}
\mathbf{XE}:=[\mathbf{(X}_{0}+\epsilon \mathbf{X}_{1}+\epsilon ^{2}\mathbf{X}%
_{2}+\cdots +\epsilon ^{k} \mathbf{X}_{k})(\mathbf{E}%
_{0}+\epsilon \mathbf{E}_{1}+\epsilon ^{2}\mathbf{E}_{2}+\cdots +\epsilon ^{k}%
\mathbf{E}_{k})]_{(\ref{5})} =O(\epsilon ^{k+1})  \!\!\!\label{7}
\end{gather}%
is satisf\/ied \cite{a4,a16,a26}. Here $\mathbf{E}_{0}$ is the exact
system of equations, $\mathbf{E}_{1}$ is the f\/irst-order approximate
part and $\mathbf{E}_{2}$ is the second-order approximate part of
the perturbed system and so on; $\mathbf{X}_{0}$~is the exact
symmetry generator, $\mathbf{X}_{1}$ the f\/irst-order approximate
part, $\mathbf{X}_{2}$ the second-order approximate part of the
symmetry generator and so on. These approximate symmetries do not
necessarily form a Lie algebra but do form a so-called
``approximate Lie algebra''~\cite{a27}.

If \eqref{5} admits \eqref{6} with $\mathbf{X}_{0}\neq 0$, then $\mathbf{X}_{0}$
is an exact symmetry of the unperturbed $\mathbf{E}_{0}=0$ and
$\mathbf{X}_{0}$~is a stable symmetry for a given perturbation. The
perturbed equation \eqref{5} always has the approximate symmetry $\epsilon
\mathbf{X}_{0}$ which is known as a trivial symmetry and
$\mathbf{X}$ given by \eqref{6} with  $\mathbf{X}_{0}\neq 0$ and
$\mathbf{X}_{1}\neq \mathbf{X}_{0}$ is called a non-trivial
approximate symmetry~\cite{a28}.

An alternate method for def\/ining approximate symmetries of DEs was
given by Fushchich and Shtelen \cite{c11}. They interchange the
order of approximation and take the limit, between the parameter
of the symmetry generator of the algebra on one hand and the
approximation parameter on the other hand. This method is compared
with that of Baikov et al.\ in~\cite {d11,e11}. For our
purpose it makes no dif\/ference which of these methods is used. We
will follow the method of Baikov et al.~\cite{b11}.

\subsection{Symmetries and approximate symmetries of the Schwarzschild
metric}

Minkowski spacetime has the Poincar\'{e} symmetry algebra
$so(1,3)\oplus _{s}\R^{4}$ (where $\oplus _{s}$ denotes semi direct
sum) with the $10$ generators
\cite{a17,a18}%
\begin{gather}
\mathbf{X}_{0}=\frac{\partial }{\partial t},\qquad
\mathbf{X}_{1}=\cos \phi
\frac{\partial }{\partial \theta }-\cot \theta \sin \phi \frac{\partial }{\partial \phi },  \label{8} \\
\mathbf{X}_{2}=\sin \phi \frac{\partial }{\partial \theta }+\cot
\theta \cos \phi \frac{\partial }{\partial \phi },\qquad
\mathbf{X}_{3}=\frac{\partial
}{\partial \phi },  \label{9} \\
\mathbf{X}_{4}=\sin \theta \cos \phi \frac{\partial }{\partial
r}+\frac{\cos \theta \cos \phi }{r}\frac{\partial }{\partial \theta
}-\frac{\csc \theta
\sin \phi }{r}\frac{\partial }{\partial \phi },  \label{10} \\
\mathbf{X}_{5}=\sin \theta \sin \phi \frac{\partial }{\partial
r}+\frac{\cos \theta \sin \phi }{r}\frac{\partial }{\partial \theta
}+\frac{\csc \theta
\cos \phi }{r}\frac{\partial }{\partial \phi },  \label{11} \\
\mathbf{X}_{6}=\cos \theta \frac{\partial }{\partial r}-\frac{\sin \theta }{r%
}\frac{\partial }{\partial \theta },  \label{12}\\
\mathbf{X}_{7}=\frac{r\sin \theta \cos \phi }{c}\frac{\partial }{\partial t}%
+ct\left(\sin \theta \cos \phi \frac{\partial }{\partial r}+\frac{\cos
\theta \cos \phi }{r}\frac{\partial }{\partial \theta }-\frac{\csc
\theta \sin \phi
}{r}\frac{\partial }{\partial \phi }\right),  \label{13} \\
\mathbf{X}_{8}=\frac{r\sin \theta \sin \phi }{c}\frac{\partial
}{\partial t} +ct\left(\sin \theta \sin \phi \frac{\partial }{\partial
r}+\frac{\cos \theta \sin \phi }{r}\frac{\partial }{\partial \theta
}+\frac{\csc \theta \cos \phi
}{r}\frac{\partial }{\partial \phi }\right),  \label{14} \\
\mathbf{X}_{9}=\frac{r\cos \theta }{c}\frac{\partial }{\partial
t}+ct\left(\cos \theta \frac{\partial }{\partial r}-\frac{\sin \theta
}{r}\frac{\partial }{\partial \theta }\right),  \label{15}
\end{gather}
$\mathbf{X}_{0}$, $\mathbf{X}_{4}$, $\mathbf{X}_{5}$ and
$\mathbf{X}_{6}$ are the spacetime translations (with a Lie
sub-algebra ($\R^{4}$)) which provide the laws of conservation of
energy and linear momentum, $\mathbf{X}_{1}$, $\mathbf{X}_{2}$ and
$\mathbf{X}_{3}$ are the rotations (with a Lie sub-algebra
($so(3)$)) which provide the laws of conservation of angular
momentum and $\mathbf{X}_{7}$, $\mathbf{X}_{8}$ and $\mathbf{X}_{9}$
are the Lorentz transformations (with a Lie sub-algebra ($so(3)$))
which provide the laws of conservation of spin angular momentum;
via Noether's theorem \cite{a2}. The symmetry algebra for the
geodesic equations is $sl(6,\R)$, which has many symmetries that do
not correspond to conservation laws \cite{a13}.

For the Schwarzschild metric, corresponding to a point mass, $M$,
situated at the origin only the f\/irst $4$ generators ($\mathbf{X}_{0},
\mathbf{X}_{1},\mathbf{X}_{2},\mathbf{X}_{3}$) apply, yielding only conservation of
energy and angular momentum. The generators
$\mathbf{X}_{5},\dots,\mathbf{X}_{9}$, yielding conservation of linear
momentum and spin angular momentum are ``lost''. Using the approximate symmetries
for the system of geodesic equations for this metric with $\epsilon =
2GM/c^2$~\cite{a1} (where $G$ is Newton's gravitational constant
and~$c$ is the speed of light in vacuum) the ``lost'' symmetries are recovered as trivial
approximate symmetries yielding approximate conservation of those
quantities. The symmetry algebra of the geodesic equations for the
Schwarzschild metric is $so(3)\oplus \R \oplus d_{2}$ (where $d_{2}$
is the dilation algebra with generators $\partial/ \partial
s$ and $s\partial /\partial s$) \cite{a13}. The geodesic
equations can be reduced to a single orbital equation that has the
two non-trivial approximate symmetries given by the generators
\begin{gather}
\begin{gathered}
\mathbf{X}_{a1} =\sin \phi \frac{\partial }{\partial u}+\epsilon
\left(2\sin\phi \frac{\partial }{\partial \phi }+u\cos \phi \frac{\partial }
{\partial u}\right), \\ 
\mathbf{X}_{a2} =\cos \phi \frac{\partial }{\partial u}-\epsilon
\left(2\cos\phi \frac{\partial }{\partial \phi }-u\sin \phi \frac{\partial }
{\partial u}\right).
\end{gathered} \label{17}
\end{gather}

\section[Symmetries and approximate symmetries of the Reissner-Nordstr\"om metric]{Symmetries and approximate symmetries\\ of the Reissner--Nordstr\"om metric}

The f\/ield of a point massive electric charge at rest at the origin
is given by the RN metric (see for example \cite{a29})
\begin{gather*}
ds^{2}=e^{\nu }dt^{2}-e^{-\nu }dr^{2}-r^{2}d\theta ^{2}-r^{2}\sin
^{2}\theta d\phi ^{2},  
\end{gather*}%
with
\begin{gather}
e^{\nu }=1-\frac{2GM}{c^{2}r}+\frac{GQ^{2}}{c^{4}r^{2}},  \label{19}
\end{gather}%
where $Q$ is the electric charge of the point gravitational source.
Electromagnetism is the only long range force in Nature other than
gravity and this is the only spherically symmetric, static exact
solution of the ``sourceless'' Einstein--Maxwell equations. In the
chargeless case ($Q=0$) it reduces to the Schwarzschild metric. It
is of interest to look at the symmetry structure of this metric and
the corresponding symmetries, and approximate symmetries of the
geodesic equations.

It had been pointed out \cite{a1} that there is a dif\/ference between
the conservation laws obtained for the system of geodesic equations
and the single, orbital equation for the Schwarzschild metric. It
was further remarked that it should be checked if this dif\/ference
also holds for other spacetimes. We investigate this question for
the orbital equation in the RN
metric,%
\begin{gather*}
\frac{d^{2}u}{d\phi ^{2}}+u=\frac{GM}{h^{2}}-\frac{GQ^{2}}{c^{2}h^{2}}u+
\frac{3GM}{c^{2}}u^{2}-\frac{2GQ^{2}}{c^{2}}u^{3},  
\end{gather*}%
where $h$ is the classical angular momentum per unit mass and $u=1/r$.
In the classical limit $c\rightarrow \infty $ it gives the classical
orbital equation and for $Q=0$ it yields the Schwarzschild orbital
equation. For the approximate symmetries of this equation we take
\begin{gather}
\epsilon =\frac{2GM}{c^{2}},\qquad \frac{GQ^{2}}{c^{4}}=k\epsilon ^{2}.
\label{21}
\end{gather}
For an RN black hole (see for example \cite{a30}) that is
\begin{gather*}
M^{2}\geq Q^{2},\qquad {\rm we \ \  have}\quad 0<k\leq1/4.
\end{gather*}%
If $k>1/4$ the metric represents a naked singularity. Taking $\epsilon =0$
in the orbital equation, the exact symmetry generators are
\begin{gather*}
\mathbf{X}_{0} =u\frac{\partial }{\partial u } , \qquad
\mathbf{X}_{1}=\cos
\phi \frac{\partial }{\partial u},\qquad \mathbf{X}_{2}=\sin \phi \frac{%
\partial }{\partial u},\qquad \mathbf{X}_{3}=\frac{\partial }{\partial \phi
},  \\ 
\mathbf{X}_{4} =\cos 2\phi \frac{\partial }{\partial \phi }-u\sin
2\phi
\frac{\partial }{\partial u},\qquad \mathbf{X}_{5}=\sin 2\phi \frac{\partial }{\partial \phi }+u\cos 2\phi \frac{\partial }{\partial u},\\
\mathbf{X}_{6} =u\cos \phi \frac{\partial }{\partial \phi
}-u^{2}\sin \phi
\frac{\partial }{\partial u},\qquad \mathbf{X}_{7}=u\sin \phi \frac{\partial }{\partial \phi }+u^{2}\cos \phi \frac{\partial }{\partial u}.
\end{gather*}
Retaining terms of f\/irst-order in $\epsilon $ and neglecting
$O(\epsilon ^{2}),$ the f\/irst approximate symmetry genera\-tors are
given by \eqref{17}. In the second approximation, that is when
we retain terms quadratic in $\epsilon $, this equation possesses
\textit{no} non-trivial second-order approximate symmetry generators,
but the f\/irst-order approximate symmetry generators are still retained.
Thus there is no new approximate conservation laws but only the previous
conservation laws that have been recovered.

A better idea of what is actually required comes from the full
system of geodesic equations. The geodesic equations are given by
\begin{gather}
\ddot{t}+\nu ^{\prime }\dot{t}\dot{r}=0, \nonumber\\ 
 \ddot{r}+\frac{1}{2}(e^{\nu })^{\prime }(e^{\nu }c^{2}\dot{t}
^{2}-e^{-\nu }\dot{r}^{2})-re^{\nu }(\dot{\theta}^{2}+\sin
^{2}\theta\dot{\phi}^{2})=0,\nonumber\\ 
 \ddot{\theta}+\frac{2}{r}\dot{r}\dot{\theta}-\sin
\theta \cos \theta\dot{\phi}^{2}=0,\nonumber\\ 
 \ddot{\phi}+\frac{2}{r}\dot{r}\dot{\phi}+2\cot\theta\dot{\theta}
\dot{\phi}=0,
 \label{29}
\end{gather}
with $e^{\nu }$ now given by \eqref{19} and hence
\begin{gather*}
\nu ^{\prime }=\frac{\epsilon }{r^{2}}+\frac{1-2k}{r^{3}}\epsilon
^{2},\qquad (e^{\nu })^{\prime }=\frac{\epsilon }{r^{2}}-\frac{2k}{r^{3}}
\epsilon ^{2}, \qquad e^{-\nu }=1+\frac{\epsilon }{r}+\frac{(1-k)}{r^{2}}
\epsilon ^{2}, 
\end{gather*}
where ``$\cdot$'' denotes dif\/ferentiation with respect to the geodetic
parameter~$s$.

To construct the determining equations for the second-order
approximate symmetries we use~\eqref{8} and \eqref{9} as the $4$ exact symmetry
generators and \eqref{10}--\eqref{15} as the $6$ f\/irst-order approximate
symmetry generators. Of the $4$ exact generators $2$ do not appear
in the new determining equations and the other $2$ cancel out. The
$6$ generators of the f\/irst-order approximate symmetry have to be
eliminated for consistency of the new determining equations, making
them homogeneous. The resulting system is the same as for Minkowski
spacetime, yielding $10$ second-order approximate symmetry
generators. Four of them are again the exact symmetry generators
used earlier, and hence simply ``add
into'' them, making no dif\/ference. The other $6$
replace the lost f\/irst-order approximate symmetry generators. The
full set has the Poincar\'{e} algebra $so(1,3)\oplus _{s} \R^{4}$
apart from some non-Noether symmetries. There are no non-trivial
second-order approximate symmetries as was the case for the
f\/irst-order approximate symmetries.

It is worth remarking that for the f\/irst-order approximate
symmetries it did not matter whether we used the full system \eqref{5} or
the un-perturbed system $\mathbf{E}_{0},$ in \eqref{7}. However, for the
second-order approximate symmetries it \textit{does} make a
dif\/ference. One needs to use the full system~\eqref{5} and \textit{not} the
un-perturbed system $\mathbf{E}_{0},$ in \eqref{7} to obtain the solution.

The exact symmetry generators include not only \eqref{8} and \eqref{9}, but also
the generators of the dilation algebra, $\partial /\partial
s$, $s\partial /\partial s$ corresponding to
\begin{gather*}
\xi (s)=c_{0}s+c_{1}.  
\end{gather*}%
In the determining equations for the f\/irst-order approximate
symmetries the terms involving $\xi _{s}=c_{0}$ cancel out. However,
for the second-order approximate symmetries the terms in $\xi _{s}$
do \textit{not} automatically cancel out but collect a scaling
factor of $(1-2k)$ so as to cancel out. (This factor comes from the
application of the perturbed system, rather than the un-perturbed
one.) Since energy conservation comes from time translational
invariance and $\xi $ is the coef\/f\/icient of
$\partial /\partial s$ in the point transformations given by
\eqref{2}, where $s$ is the proper time, the scaling factor $(1-2k)$
corresponds to a re-scaling of energy. Thus, whereas there was no
energy re-scaling needed for the f\/irst-order approximate symmetries,
it arises naturally in the second-order approximate symmetry. Using
\eqref{21} we get the energy re-scaling factor
\begin{gather*}
(1-2k)=(1-Q^2/2GM^2). 
\end{gather*}
Thus, even though there are no non-trivial second-order approximate
symmetries, we get the non-trivial result of energy re-scaling from
the second-order approximation. This point will be discussed further
in the next section.

It is worth remarking that when some symmetries are
``lost'' at one order (exact or
f\/irst-order approximate) they are ``recovered'' at the next (at least to second-order)
as ``trivial'' approximate
symmetries.

\section{Summary and discussion}
We studied the approximate symmetries of the RN metric. This metric
has isometry algebra $so(3)\oplus \R$ with generators \eqref{8} and \eqref{9}. The
symmetry algebra of the geodesic equations for this metric is
$so(3)\oplus \R \oplus d_{2}$. We used Lie symmetry methods for DEs
to explore the second-order approximate symmetries of the RN metric.
Neglecting terms containing $\epsilon ^{2}$ in the geodesic
equations 
\eqref{29} this metric has the same f\/irst approximate
symmetries as those of the Schwarzschild metric. Again we get
\textit{no} non-trivial approximate symmetry generator in the second
approximation. We only recover the ``lost'' conservation laws as approximate
conservation laws. As for the Schwarzschild metric, where there is a
dif\/ference between the conservation laws obtained for the system of
geodesic equations and for the single orbital equation, the
dif\/ference also holds for the RN metric.

The re-scaling of energy for the RN metric, which does not appear
for the Schwarzschild metric is of special interest. Notice that the
pseudo-Newtonian formalism \cite{a31,a32,a33,a34} gives re-scaling
of force by $(1-Q^2/rMc^2)$. The reduction is by the
ratio of the electromagnetic potential energy at a distance $r$ to
the rest energy of the gravitational source. It is position
dependent. That would not be reasonable for the energy in the f\/ield
by itself. The scaling obtained here, $(1-Q^2/2GM^2)$, is
more reasonable as relating the electromagnetic self-energy to the
gravitational self-energy, with the radial parameter, $r$, canceled
out.

It would be of interest to apply the def\/inition of second-order
approximate symmetries to the Kerr metric where there are only two
isometries $\mathbf{X}_{0}=\partial /\partial t$, $\mathbf{X}_{3}=\partial /\partial \phi $ and a non-trivial
\textit{Killing tensor}~\cite{a18}\ to see if the same result of
energy re-scaling holds. It would also be important to check if the
energy in the gravitational f\/ield could be obtained by considering
approximate symmetries of gravitational wave spacetimes. For this
purpose one can f\/irst consider some non-f\/lat static spacetime. Since
gravitational wave spacetimes are non-static solutions of the vacuum
Einstein f\/ield equations (EFEs), one should perturb the static
spacetime with a time dependent small parameter and then look at the
approximate symmetries of this gravitational wave-like spacetime. As
gravitational wave spacetimes are solutions of vacuum EFEs, one
would have to calculate the Weyl and stress-energy tensor for this
gravitational wave-like spacetime (for which the stress-energy
tensor is non-zero) to equal orders of approximation to see how much
energy is contained in the gravitational f\/ield and how much in the
matter f\/ield. To understand the actual problem one will then have
consider an exact gravitational wave solution and look at its
approximate symmetries. For the exact gravitational wave solutions a
formula for the momentum imported to test particle is already
available~\cite{a35}. The comparison of the two results might enable
one to identify a physically signif\/icant energy content in the
gravitational wave spacetime.

\subsection*{Acknowledgements}

IH would like to thank Pakistan Science Foundation (PSF) for their
full f\/inancial support to attend the Seventh International
Conference ``Symmetry in Nonlinear Mathematical Physics'', where this
work was presented. AQ would like to thank the conference
organizers for the local hospitality and International Mathematical
Union (IMU) for the traveling grant and DECMA and the School for
Computational and Applied Mathematics of the University of
Witwatersrand, Johannesberg, where the writing-up was completed.

\pdfbookmark[1]{References}{ref}
\LastPageEnding

\end{document}